\documentclass[pra,twocolumn,superscriptaddress,superscriptaddress,amsmath]{revtex4-2}

\usepackage{graphicx}
\usepackage{float}
\usepackage{epstopdf,cancel}
\usepackage{epsf,latexsym,bbm}
\usepackage[mathscr]{euscript}

\usepackage{amssymb,amsmath}
\usepackage{mathtools} 
\usepackage{times,graphics}
\usepackage{soul,xcolor}
 \usepackage{epsfig,ulem}

\def\6{\langle}
\def\9{\rangle}

\newcommand{\etal}{\textit{et al.}}

\newcommand{\defeq}{\vcentcolon=}
\newcommand{\eqdef}{=\vcentcolon}

\usepackage{url,hyperref}
\hypersetup{colorlinks,linkcolor={blue!55!black},citecolor={red!45!black},urlcolor={blue!45!black},breaklinks=true}

\begin{document}

\title{Light propagation in Kerr spacetime}

\author{Pravin Kumar Dahal}
\email{pravin-kumar.dahal@hdr.mq.edu.au}
\affiliation{School of Mathematical \& Physical Sciences, Macquarie University}

\date{\today}

\begin{abstract}
    We explicitly solve the equations for the propagation of an electromagnetic wave up to the subleading order geometric optics expansion in the Kerr spacetime. This is done in two nontrivial steps. We first construct a set of parallel propagated null tetrad in Kerr spacetime. Two of the components of such tetrad give the propagation and polarization of an electromagnetic wave in geometric optics approximation. Then we use the parallel propagated tetrad to solve the modified trajectory equation in Kerr spacetime. We obtain the wavelength-dependent deviation of the trajectory of electromagnetic waves, which gives the mathematical description of the gravitational spin Hall effect in Kerr spacetime.
\end{abstract}

\maketitle

\section{Introduction}

One way to study a black hole is from the light emitted by the matter in the vicinity of the black hole. Such radiation has to propagate a long distance through the strong gravitational fields before reaching the observer. So, one needs to solve the equation for the propagation of electromagnetic waves in curved spacetime in order to extract the information contained in astrophysical observations.

Solution of the wave equations, in general, in curved spacetimes is a cumbersome task. A well-known and widely employed approximation technique developed long ago for this purpose is geometric optics approximation~\cite{mtw7, d35}. In this approximation, valid in the limit of an infinite frequency, we reduce the wave equation to the ray equation and transport equation along the ray. However, the geometric optics approximation does not capture the wavelike behaviour, which is essential for the wave of large but finite frequency propagating in curved spacetime. For this purpose, geometric optics approximation is generalized by assuming it as the leading order approximation of some perturbative expansion of the trajectory and transport equation in terms of the inverse frequency $1/\omega$ of the wave~\cite{f31,38,39}. We call spin optics to this approximation up to the subleading order in $1/\omega$, and it is sufficient to describe physical phenomena originating from the interaction of spin with the extrinsic orbital angular momentum of the wave~\cite{f31,14,39}. An example of such a phenomenon we discuss here is the gravitational spin Hall effect~\cite{opj8}. This article is concerned only with extrinsic and not intrinsic angular momentum.

The analogous phenomenon known as the spin Hall effect or optical Magnus effect is observed in condensed matter physics~\cite{17,18,19}. This effect results when light propagates in an inhomogeneous medium, where the inhomogeneous medium imparts orbital angular momentum, thereby interacting with the intrinsic spin angular momentum of light~\cite{brn15}. This effect is well developed~\cite{45} and experimentally verified~\cite{wks17,kmg18,18,19}, thus serving as a basis for our calculation and comparison.

Modified geometric optics was initially developed for stationary spacetimes in Refs~\cite{d35,f31,39}. Recently, it was devised for general spacetimes in Refs.~\cite{13,14}. We have also developed the covariant formulation of spin optics in general spacetime in Ref.~\cite{pkd1,pkd2}, which yield slightly different results. Using the tools developed there, we demonstrate the gravitational spin Hall effect in the Kerr background analytically and numerically. Previously, Ref.~\cite{fs46} has used the subleading order geometric optics correction, devised for the stationary spacetimes, to demonstrate the gravitational spin Hall effect in the Kerr spacetime (see, also, Ref.~\cite{ms47}).
Electromagnetic waves from astrophysical sources might pass through the vicinity of massive gravitating bodies, which act as gravitational lenses. The leading order geometric optics expansion suffices if the characteristic wavelength of electromagnetic waves is negligible compared to the length scale of the inhomogeneities in spacetime curvature from such bodies. However, subleading order correction from the geometric optics might be necessary when studying gravitational lensing of electromagnetic waves with wavelengths larger than the Schwarzschild radius of lensing objects~\cite[]{24,25}. In such a situation, wave effects must be appropriately taken into account. To perform this task, we proceed in two steps. First, we construct the parallel propagated null tetrad, which is the solution of the wave equation in the geometric optics approximation. Then, we need to use that tetrad to find the subleading order correction in the trajectory. Propagation and polarization relations in the subleading order are calculated explicitly in terms of this tetrad.

This article is organized as follows. In Sec.~\ref{ppnt}, we describe the general procedure of solving the parallel transport equations for null geodesics in Petrov type-D spacetimes admitting an additional integral of motion~\cite{sc21}, which is quadratic in particle momenta. We then apply this procedure to find the parallel propagated null tetrad in Kerr spacetime. In Sec.~\ref{tdshe}, we use this parallel propagated tetrad to find the leading order deviation in the null trajectory from the geometric optics approximation analytically and numerically. We then discuss our results and conclude the article in Sec.~\ref{dac}.

We consider a spacetime manifold $M$ with the metric $g_{\mu\nu}$ of Lorentzian signature $\left(-,+,+,+\right)$. The phase space is the cotangent bundle $T^* M$, whose points are written as $(x, p)$. Similarly, $\tilde m$ represents the complex conjugate of $m$. We use the system of units with $G=c=1$ and adopt the Einstein summation convention. A semicolon ($;$) denotes the covariant derivative, $\lambda$ denotes the parameter of electromagnetic wave curves and $\dot x= d x/d\lambda$. The curvature convention is adopted from Ref.~\cite{mtw7}.

\section{Parallel propagated null tetrad} \label{ppnt}

In general, it is not straightforward to explicitly solve the parallel transport equations. Fortunately, the Kerr geometry has some remarkable properties, including the separability of wave equations and the complete integrability of geodesics. These special separability properties that permit the explicit integration of the geodesic equations can also be used to provide an explicit solution to the problem of constructing a parallel propagated orthonormal tetrad. For timelike geodesics, this was done by Marck~\cite{mk3} in two steps:
1) First, construct a locally defined orthonormal tetrad along a null geodesic such that two of its components are already parallel propagated.
2) Then obtain the entirely parallel transported tetrad by rotating along some spatial hypersurface. The rotation angle would be the sum of two functions, one of $r$ coordinates only and the other of $\theta$ coordinates expressed in terms of elliptic integrals.

The procedure would be different for the null geodesic. Unlike in this reference, we will not restrict ourselves to the static orthonormal frame and will solve the equations of parallel transport in general. This parallel propagated tetrad would be useful in studying tidal effects near the Kerr black hole and the gravitational spin Hall effect. They are also useful while investigating particles and fields with spins~\cite{sc17}. In quantum physics, the point-splitting method is used to calculate the renormalized values of local observables in curved spacetime, and it relies on the parallel transported frame. It also plays a role in proving the peeling-off property of the gravitational radiation~\cite{rs4,rs5,np6}.

Using the separation of variables on the Hamilton-Jacobi equation, Carter found the existence of the fourth constant of the Kerr spacetime, making the geodesic equation analytically soluble~\cite{bk8}. This constant of motion is quadratic in particle momenta~\cite{wp9} and has a direct correspondence with the rank two Killing tensor. A rank two Killing tensor $K_{\alpha\beta}$ is a symmetric tensor with vanishing symmetrized covariant derivative, $\nabla^{(\gamma} K^{\alpha\beta)}=0$.  The spacetime symmetry associated with the Killing tensor of rank two and higher is known as hidden symmetry.

Spacetime with some symmetries always has its counterpart in the phase space. However, not all phase space symmetries are associated with the configuration space. Symmetries that can be reduced to the configuration space are explicit symmetries of the spacetime, and those which do not have its counterpart in the configuration space are hidden symmetries. Explicit continuous symmetries are described by the Killing vectors and hidden symmetries by the Killing tensors. Killing tensors do not generate a spacetime diffeomorphism; hence, they are not associated with the configuration space. However, the existence of these tensors could be realized from the geodesic equations of motion.

Following Carter's discovery, Penrose~\cite{rp10} and Floyd~\cite{rf11} showed that there exists the Killing-Yano tensor $f_{\alpha\beta}$ in the Kerr geometry, which resembles a square root of the Killing tensor and obeys $\nabla_{(\gamma} f_{\alpha) \beta}=0$. Its Hodge dual $h_{\alpha\beta}= *f_{\alpha\beta}$ is again a two-form that satisfies
\begin{equation}
    \nabla_\gamma h_{\alpha\beta}= g_{\gamma\alpha} \xi_\beta- g_{\gamma\beta} \xi_\alpha; \quad \xi_\alpha= \frac{1}{D-1} \nabla_\beta h^\beta_{~~\alpha}, \label{eq13}
\end{equation}
where $\xi_\alpha$ is the Killing vector and $D$ is the spacetime dimension (for our case, $D=4$). The object satisfying such an equation is a closed conformal Killing-Yano two-form. The wedge product of these two-forms is also a closed conformal Killing-Yano tensor~\cite{pk12}. This quantity is called a principal tensor by Frolov \etal~\cite{vf13}. The principal tensor describes hidden symmetry of the spacetime, whose existence ensures the complete integrability of geodesic motion (consequences extend beyond this property; see Ref.~\cite{vf13}). We will show below that the principal tensor $h_{\alpha\beta}$ also enables us to construct a complete set of parallel transported frames along the geodesics. In summary, the integrability of the geodesic equations is ensured by the existence of Killing tensor $K_{\alpha\beta}$, whose square root resembles the Killing-Yano two-form $f_{\alpha\beta}$. As a result, we could also analytically solve the equations of parallel transport applied to an orthonormal tetrad along a null geodesic congruence.

\subsection{Parallel transport along null geodesics}

We consider an affine parametrized null geodesics ${\cal C}$, with tangent vector $l^\alpha$. Let $h^{\alpha\beta}$ be the principal tensor. Then, defining
\begin{equation}
    v^\alpha= h^{\alpha\beta} l_\beta- A(r,\theta) l^\alpha, \label{eq18}
\end{equation}
we obtain
\begin{equation}
    \dot v^\alpha= l^\alpha \left(-\dot A+ l^\beta \xi_\beta\right),
\end{equation}
where Eq.~\eqref{eq13} has been used to find this. Thus, the requirement that $v^\alpha$ be parallel transported gives
\begin{equation}
    \dot A= l^\beta \xi_\beta. \label{da4}
\end{equation}
Now, using $v^\alpha$ as a seed vector and defining the equation analogous to Eq.~\eqref{eq18} enables us to construct another parallel transported vector immediately
\begin{equation}
    n^\alpha= h^{\alpha\beta} v_\beta- B(r,\theta) l^\alpha, \qquad \dot B= v^\beta \xi_\beta. \label{eq21}
\end{equation}
The vector $n^\alpha$ does not belong to the null plane of vectors orthogonal to $l^\alpha$, because of which it cannot be used as a new seed to construct equations like Eq. (2) to create new parallel transported vectors.

We can define another parallel transported vector depending on particle momentum $l^\alpha$ and position through the tensor $f_{\alpha \beta}$
\begin{equation}
    u^\alpha= f^\alpha_{~~\beta} l^\beta, \label{eq22}
\end{equation}
Then, this quantity satisfies the parallel transport equations if and only if tensor $f_{\alpha\beta}$ satisfies the Killing tensor equation~\cite{ch14}
\begin{equation}
    \nabla_{(\alpha} f_{\beta\gamma)}= 0.
\end{equation}
If the tensor is completely antisymmetric, it is called the Killing-Yano form~\cite{ky15}
\begin{equation}
    \nabla_{(\alpha} f_{\beta)\gamma}= 0.
\end{equation}
As $f_{\alpha \beta}= *h_{\alpha \beta}$ is the Hodge dual of $h_{\alpha \beta}$, the above two equations indeed hold, and hence $f_{\alpha\beta}$ is parallel transported along the geodesics. This is because the Hodge dual of a closed conformal Killing-Yano tensor is a Killing-Yano tensor and vice versa. Corresponding vector $u^\alpha$ is parallel propagated and also perpendicular to the momentum of the particle
\begin{equation}
    u_\alpha l^\alpha=0.
\end{equation}
Conversely, any skew-symmetric vector $u^\alpha$ that is linear in momentum and parallel propagated along and orthogonal to any geodesic describes the Killing-Yano tensor $f_{\alpha\beta}$. Thus, by construction, vectors $(l^\alpha, n^\alpha, v^\alpha, u^\alpha)$ are parallel transported along the geodesic ${\cal C}$. This property has been used to explicitly construct the parallel propagated frame in Kerr spacetime~\cite{mk3,mk16} and higher dimensions~\cite{pc18}.

\subsection{Application to Kerr spacetime}

Let us take the Kerr metric in Boyer-Lindquist coordinates
\begin{multline}
    ds^2= -\left(1- \frac{2 M r}{\rho^2}\right) dt^2- \frac{4 a M r}{\rho^2} \sin^2\theta dt d\phi+ \frac{\rho^2}{\Delta} dr^2+ \\ \rho^2 d\theta^2+ \frac{{\cal A}}{\rho^2} \sin^2\theta d\phi^2,
\end{multline}
where
\begin{align}
    \rho^2=& r^2+ a^2 \cos^2\theta, \quad \Delta= r^2+ a^2- 2 M r,\nonumber\\
    {\cal A}=& (r^2+ a^2)^2- \Delta a^2 \sin^2\theta.
\end{align}
The equation of motions for null geodesics in Kerr spacetime is
\begin{align}
    \dot t=& \frac{{\cal A}- 2 a M r \zeta}{\Delta \rho^2},\quad
    \rho^4 \dot r^2= {\cal R}^2,\nonumber\\
    \rho^4 \dot\theta^2=& \Theta^2,\quad
    \dot \phi= \frac{2 a M r+ (\rho^2- 2 M r) \zeta \csc^2\theta}{\Delta\rho^2},
\end{align}
where $\eta$, $\zeta$ are constants of motion for the null geodesic trajectory and
\begin{align}
    {\cal R}^2=& \left(a^2-a \zeta +r^2\right)^2-\Delta \left((a-\zeta )^2+\eta \right),\nonumber\\
    \Theta^2=& a^2 \cos ^2\theta+\eta -\zeta ^2 \cot ^2\theta.
\end{align}
One of the obvious choices of the parallel propagated orthonormal tetrad would be the unit vector tangent to the geodesic
\begin{equation}
    l_{0 \mu}= \left(-1, \frac{{\cal R}}{\Delta}, \Theta, \zeta\right). \label{im14}
\end{equation}
Now, to calculate three others, we take the principal tensor of Kerr spacetime in coordinates $(t,r,\theta,\phi)$

\begin{widetext}
\begin{equation}
    h_{\alpha\beta}= \left(
\begin{array}{cccc}
 0 & r & a^2 \sin \theta \cos \theta & 0 \\
 -r & 0 & 0 & a r \sin ^2\theta \\
 -a^2 \sin \theta \cos \theta & 0 & 0 & a \left(a^2+r^2\right) \sin \theta \cos \theta \\
 0 & -a r \sin ^2\theta & -a \left(a^2+r^2\right) \sin \theta \cos \theta & 0 \\
\end{array}
\right),
\end{equation}
and the corresponding Killing vector $\xi^\mu= (1,0,0,0)$ related by Eq.~\eqref{eq13}. This allows us to calculate one of the parallel transported vectors using Eq.~\eqref{eq18}
\begin{multline}
    v_\alpha= \bigg(\frac{A(r,\theta) \rho^2+ r {\cal R}+a^2 \sin \theta \cos \theta \Theta}{\rho^2},-\frac{A(r,\theta) {\cal R}+a^2 r-a \zeta  r+r^3}{\Delta},a (\zeta  \cot \theta-a \sin \theta \cos \theta)-A(r,\theta) \Theta,\\
    -\frac{\zeta  A(r,\theta) \rho^2+a r \sin ^2\theta {\cal R}+a \left(a^2+r^2\right) \sin \theta \cos \theta \Theta}{\rho^2}\bigg).
\end{multline}
Moreover, using Eq.~\eqref{da4} for the derivative of $A(r,\theta)$, we could also calculate $A$ as
\begin{equation}
    \dot A(r,\theta)= -1, \qquad \mathrm{or} \quad A(r,\theta)= {\cal F}(r)+ {\cal G}(\theta), \label{eq24}
\end{equation}
where
\begin{equation}
    {\cal F}(r)= -\int \frac{r^2}{{\cal R}} dr; \quad {\cal G}(\theta)= -\int \frac{a^2 \cos^2\theta}{\Theta} d\theta.
\end{equation}Now, we have all the ingredients that allow us to integrate $\dot B$ appearing in Eq.~\eqref{eq21}, from which one can express $B$ in terms of $A$ as
\begin{equation}
    B(r,\theta)= \frac{1}{2} \left(r^2-a^2 \cos ^2\theta-A^2(r,\theta)\right).
\end{equation}
Thus, using Eq.~\eqref{eq21}, we could calculate another parallel transported vector
\begin{multline}
    n_{0\alpha}= \bigg(\frac{-4 A(r,\theta) \left(2 r {\cal R}+a^2 \sin 2\theta \Theta\right)-4 A^2(r,\theta) \rho^2+2\rho^2 \left(a^2 \cos 2\theta-3 a^2+4 a \zeta -2 r^2\right)}{8 \rho^2},\\
    \frac{2 A^2(r,\theta) {\cal R}+4 r \left(a^2-a \zeta +r^2\right) A(r,\theta)+2\rho^2 {\cal R}}{4 \Delta},
    \frac{1}{4} \left(2 A^2(r,\theta) \Theta+2 a A(r,\theta) (a \sin 2\theta-2 \zeta  \cot \theta)-2\rho^2 \Theta\right),\\
    \frac{8 a \sin \theta A(r,\theta) \left(r \sin \theta {\cal R}+\left(a^2+r^2\right) \cos \theta \Theta\right)+4 \zeta  A^2(r,\theta) \rho^2-4\rho^2 \left(\zeta a^2 \sin^2\theta+ \left(a^2+ r^2 \right) \left(\zeta- 2 a^2\sin^2\theta \right)\right)}{8 \rho^2}\bigg).
\end{multline}
To calculate the last parallel transported vector, we need the Hodge dual of $h_{\alpha\beta}$. It is given as
\begin{equation}
    *h_{\alpha\beta}= \left(
\begin{array}{cccc}
 0 & -a \cos \theta & a r \sin \theta & 0 \\
 a \cos \theta & 0 & 0 & -a^2 \sin ^2\theta \cos \theta \\
 -a r \sin \theta & 0 & 0 & r \left(a^2+r^2\right) \sin \theta \\
 0 & a^2 \sin ^2\theta \cos \theta & - r \left(a^2+r^2\right) \sin \theta & 0 \\
\end{array}
\right).
\end{equation}
From Eq.~\eqref{eq22}, the final parallel transported vector is
\begin{multline}
    u_\alpha= \bigg(\frac{a r \sin \theta \Theta-a \cos \theta {\cal R}}{\rho^2},\frac{a \cos \theta \left(a^2-a \zeta +r^2\right)}{\Delta},\zeta  r \csc \theta-a r \sin \theta,\\
    \frac{\sin \theta \left(a^2 \sin \theta \cos \theta {\cal R}-r \left(a^2+r^2\right) \Theta\right)}{\rho^2}\bigg).
\end{multline}
\end{widetext}

These null trajectories $l_0^\mu$ and polarization vectors $m_{0\mu}= (v_\mu+ i u_\mu)/2$ constitute the solutions of electromagnetic wave equations in the geometric optics approximation (see Ref.~\cite{14}).

\section{Transverse deflection due to the spin Hall effect} \label{tdshe}

After constructing the Fermi transported (parallel propagated) null tetrad, we study the spin Hall effect on the Kerr spacetime. Let us consider the lensing object described by the Kerr geometry with mass $M$ and angular momentum per unit mass $a$. The geometric optics approximation is valid only when the characteristic wavelength of the waves is much smaller than the horizon radius ($1/\omega \ll M+ \sqrt{M^2- a^2}$)~\cite[]{32,33}. To get the subleading order correction of the null trajectory, we substitute the above relations into the propagation equation (see Appendix \ref{soD})~\cite{pkd2}
\begin{equation}
    \frac{D^2 x^\alpha}{D\lambda^2}= -\frac{i}{\omega} R^\alpha_{~\beta\mu\nu} l_0^\beta m_0^\mu \tilde m_0^\nu.\label{te56}
\end{equation}
The results are cumbersome in general, but a numerical solution is possible. However, restricting ourselves near the equatorial plane considerably simplifies calculations, and we will present our results on the transverse deflection of the light ray trajectory there. At the equatorial plane $\theta= \pi/2$, $\eta= 0$ and
\begin{align}
    &\frac{D^2 t}{D\lambda^2}= 0, \quad \frac{D^2 r}{D\lambda^2}= 0,\nonumber\\
    &\frac{D^2 \theta}{D\lambda^2}= -\frac{3 M (a-\zeta ) A\left(r,\pi /2\right)}{\omega r^6}, \quad \frac{D^2 \phi}{D\lambda^2}= 0.\label{te77}
\end{align}
where $A\left(r,\pi /2\right)= {\cal F}(r)$. Integrating this equation gives
\begin{equation}
    \dot\theta= -\frac{3 M (a-\zeta )}{\omega r} \int \frac{{\cal F}(r)}{r^3 {\cal R}} dr \bigg|_{\eta=0}, \label{dt25}
\end{equation}
where the integration constant is chosen to be zero as the term containing it gives divergent results on integration. Thus, the deflection of light rays in the $\theta$-direction as it passes close to the lensing object of mass $M$ and angular momentum $a M$ could be calculated using the relation
\begin{equation}
    d\theta= \frac{\dot\theta}{\dot r} dr= -\frac{3 M (a-\zeta ) r}{{\omega \cal R}} \int \frac{{\cal F}(r)}{r^3 {\cal R}} dr \bigg|_{\eta=0}dr.
\end{equation}
This result for the transverse deflection is very similar to that for the Schwarzschild spacetime given in Ref~\cite{pkd1}. However, the difference comes from the appearance of the $a-\zeta$ term, instead of $\zeta$ only, which tends to cancel the magnitude of the spin Hall effect depending on the angular momentum of particle $\zeta$ and hole $a$. Explicitly,
 for the Schwarzschild case, $a=0$ and $\eta=0$
 \begin{equation}
     {\cal F}(r)= -\int \frac{r^2}{\sqrt{r^4- \left(1- \frac{2 M}{r}\right) r^2 \zeta^2}} dr.
 \end{equation}
 A photon starting from infinity and approaching the lensing object within the closest distance of $R_0$ has $\zeta \approx R_0$, up to leading order in $M/r$. We can now integrate, up to leading order in $M/r$, to obtain the total deflection angle of electromagnetic waves in the $\theta$-direction while passing near a lensing  object of mass $M$
\begin{equation}
    \theta= 2 \left(\theta_\infty -\theta_0\right)= \frac{\pi M}{2 \omega R_0^2}.
\end{equation}
This transverse deflection agrees with the result reported in Ref.~\cite[]{22} (it is also interesting to compare the magnitude of this transverse deflection with the second-order parametrized post-Newtonian expansion of the total deflection in the geometric optics approximation~\cite{eg38}). Note that this deflection is half the deflection of gravitational waves due to the spin-orbit correction~\cite[]{pkd1}.

For comparison, we take the situation where the external orbital angular momentum of the light is zero $\zeta=0$ and consider only the angular momentum of the hole. $\zeta=0$ corresponds to principle null rays with $\eta= a^2$. As we want to approximate up to the leading order expansion in $M/r$ (parameter $a$ should scale accordingly in the post-Newtonian approximation~\cite{cw44}), we could take $\eta\approx 0$. Thus, we have an electromagnetic wave propagating through an equatorial plane $\theta= \pi/2$ of a rotating lensing object. In this circumstance, we have
 \begin{equation}
     {\cal F}(r)= -\int \frac{r^2}{\sqrt{\left(r^2+ a^2\right)^2- \Delta a^2}} dr.
 \end{equation}
We could substitute this into Eq.~\eqref{dt25} to obtain $\dot\theta$. We have
\begin{equation}
     \dot\theta= -\frac{a M}{\omega r^4}.
\end{equation}
This is the only nonvanishing subleading order contribution (which could be confirmed by directly integrating Eqs.~\eqref{te77}). So, from Eq.~\eqref{tr64}, we could write subleading order components of the propagation vector as
\begin{equation}
     l_{1\mu}- b_\mu= \left(0,0, -\frac{a M}{r^2},0\right).
\end{equation}
Now, let us write $l_{1\mu}- b_\mu= \nabla\Psi$ for some $\Psi$. Then on integration, we obtain
\begin{equation}
     \Psi= -\int_{\pi/2}^\theta \frac{a M}{r^2} d\theta= -\frac{a M}{r^2} (\theta- \pi/2)\approx \frac{a M}{r^2} \cos\theta,
\end{equation}
where we have used the result that $-(\theta- \pi/2)$ is the leading order expansion of the $\cos\theta$ near $\pi/2$. Note that the gradient of this solution for $\Psi$ also contributes to the radial trajectory. However, this contribution is not on the leading order; thus, the deflection along the $\theta$ direction in the subleading order remains unchanged. As a result, the magnitude of the transverse deflection due to the spin-orbit interaction would not be affected. Thus
\begin{equation}
     \Psi= \frac{\mathbf{J}\cdot \mathbf{r}}{r^3}, \quad \mathrm{for} \quad |\mathbf{J}|= a M, \label{sp33}
\end{equation}
gives the gravitomagnetic scalar potential. In these settings, which are not manifestly covariant, the dispersion relation given in Eq.~\eqref{hj47} reduces to the form
\begin{equation}
    \mathbf{l_0}\cdot \mathbf{l_0}+\frac{2}{\omega} \nabla\Psi \cdot \mathbf{l_0} =0,
\end{equation}
which matches the dispersion relation given in Ref.~\cite{ms42,ms43}. Although the magnitude of the transverse deflection matches, other results, like the time delay, would differ from these calculations (see above Eq.~\eqref{sp33}). Also, note that deviation in trajectory caused by the angular momentum of hole $a$ is in the opposite direction to the deviation by the particle's angular momentum $\zeta$.

\subsection{Numerical results}

\begin{figure}[htbp]
\includegraphics[width=0.45\textwidth]{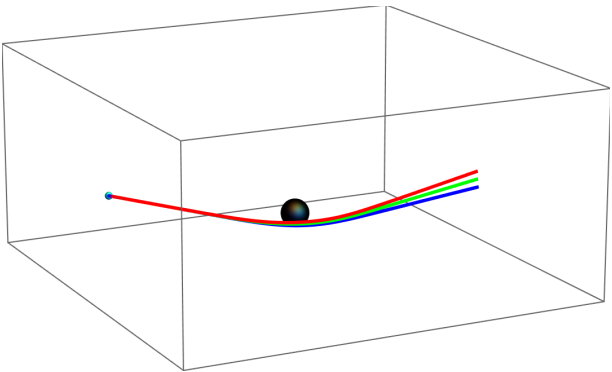}
\caption{ \label{f1} Demonstrating the spin Hall effect in Kerr spacetime for the black hole of mass $M=1$ and angular momentum per unit mass $a= 0.9$. Here, the red trajectory is for $s=+1$, the blue trajectory is for $s=-1$, and the green trajectory is for $s=0$ (geodesic curve). To make the effect visible, we have chosen an exaggerated value of the characteristic frequency $1/\omega=0.7$. The light source is on the equator $\theta= \pi/2$ with an arbitrary initial position $\phi= 7 \pi/4$ and $r=20$ and emits light at time $t=0$. The initial wave vector of these ingoing light rays is given by Eq.~\eqref{im14} with a negative sign of ${\cal R}$ and a positive sign of ${\Theta}$. This initial condition corresponds to the parameter $\lambda=0$, and we continue the propagation up to $\lambda=50$. We chose the constants of motion for the trajectory $\eta=0$ and $\zeta=50$.}
\end{figure}

Let us take the propagation Eq.~\eqref{te56}
\begin{equation}
    \frac{D^2 x^\alpha}{D\lambda^2}= -\frac{i}{\omega} R^\alpha_{~\beta\mu\nu} l^\beta m^\mu \tilde m^\nu.
\end{equation}
These are coupled second-order differential equations, and we do numerical integration and plot the results. This calculation up to the subleading order approximation is facilitated by the fact that we could substitute the leading order term for each tetrad component. Every quantity here can be computed in terms of the parallel propagated null tetrad. We thus can solve these equations using Mathematica, for which we use the default integration method, precision and accuracy. We can show that these quantities $(x(\lambda), l(\lambda))$ are gauge invariant~\cite{pkd2}, which thus makes our result gauge invariant. Fig.~\ref{f1} shows the result of our numerical simulation depicting the gravitational spin Hall effect in the Kerr background. We have exaggerated the values of the parameters to make the effect visible. The plots are for the rays of opposite circular polarization $s=\pm 1$, along with the ray of spin-zero for reference. The circularly polarized rays are not confined on the plane even at the equator, even though the null geodesics can be restricted on the plane in Kerr spacetime.

\begin{figure}[htbp]
\includegraphics[width=0.45\textwidth]{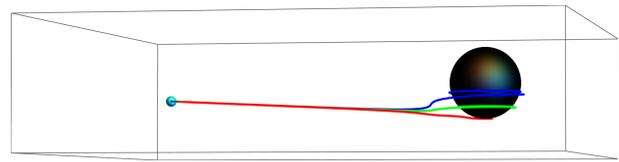}
\caption{ \label{f2} Demonstrating the spin Hall effect for the principal null rays (corresponding to $\zeta=0$ and $\eta= a^2$) in Kerr spacetime. To make the effect visible, we chose the angular momentum per unit mass $a=0.99$ and the helicity $s=\pm 2$ (positive sign for the red and negative sign for the green trajectory). We have shown in Ref.~\cite{pkd1} that propagation Eq.~\eqref{te56} holds for the gravitational waves also, which have the helicity of two after we take account of the difference in the helicity. Other values of the parameters/coordinates are precisely the same as in Fig.~\ref{f1}. Because of infinite gravitational blueshift, geometric optics approximation holds near the horizon regardless of the gravitational field strength. This assumption also plays an important role in deriving the Hawking radiation~\cite{h22}.}
\end{figure}

It is also interesting to consider the principal null rays, for which $\zeta=0$. Although such rays do not encounter the gravitational spin Hall effect in Schwarzschild spacetime, this is not true for Kerr spacetime. All null trajectories generally undergo the gravitational spin Hall effect in Kerr spacetime. The deviation for the principal null trajectories is shown in Fig.~\ref{f2}.

Observation of Fig.~\ref{f1} and \ref{f2} shows that the deviation in the trajectory caused by the angular momentum of the particle $\zeta$ is in the opposite direction to the deviation by the angular momentum of the hole $a$. Also, as the rays are scattered at a finite angle, the separation between the rays increases with increasing distance, there is no reintersection of rays. The gravitational spin Hall effect is caused by the interaction between the spin and orbital angular momentum of a particle. When the particle crosses the distance of closest approach, neither its spin nor orbital angular momentum change direction, which causes the spin-orbit interaction effect to only grow (compare Fig.~\ref{f1} with the figures in Ref.~\cite{13}). So, this effect is visible at large distances from the lensing object.

\section{Discussions and conclusion} \label{dac}

We have presented our numerical work to obtain the precise prediction of the gravitational spin Hall effect by solving the trajectory equation for the electromagnetic waves propagating in the Kerr spacetime. The trajectory equation was derived using WKB expansion, where both the phase and amplitude were expanded in inverse powers of an expansion parameter $\omega$. This expansion in both the phase and amplitude was necessary because the requirement to use the parallel propagated null tetrad in the geometric optics approximation limits the flexibility of its transformation. Generalizing to the subleading order requires using the Fermi-transported null tetrad, which also constrains its transformation properties, resulting in an observer/gauge independent spin Hall effect. This result of observer-independence was proved in Ref.~\cite{pkd2}. Thus, the spin Hall effect is distinct from observer/emitter-dependent effects (Wigner rotation and gravitational Faraday rotation) that occur in the geometric optics regime, which could be mitigated by some encoding schemes (see Ref.~\cite{pkd19}).

Additionally, the gravitational spin Hall effect results from the interaction between the spin and orbital angular momentum. Neither spin nor orbital angular momentum changes direction as light rays pass the distance of the closest approach, thus causing the spin-orbit interaction effect to grow, not decrease. As a  result, the initially divergent trajectories between the rays of different frequencies do not reconverge, which is confirmed by numerical investigations.

Although weak, it is possible to observe the gravitational spin Hall effect: first, the angular deviation, although small, can be observed at sufficiently large distances. Second, a weak quantum measurement technique could detect the spin Hall effect even if the spatial separation between the two circularly polarized light rays in opposite directions is smaller than their wavelength~\cite{hk20}.

\acknowledgments

PKD is supported by an International Macquarie University Research Excellence Scholarship.

\section*{Data Availability Statement}

Data sharing is not applicable to this article as no new data were created or analyzed in this study.

\appendix

\section{WKB approximation} \label{s3}

When the typical wavelength of the wave, propagating on approximately flat spacetime, is very small (but nonnegligible) with respect to the length scale of its amplitude and wavelength variations and the inhomogeneity of the spacetime curvature where it travels, then spin optics correction is required. We could formulate spin optics by using the WKB expansion
\begin{equation}
    A^\alpha=a^\alpha e^{i \omega {\cal S}},\label{a8}
\end{equation}
where $a^\alpha$ is the complex amplitude, and $\omega {\cal S}$ is the real phase. Here, $\omega$ is the characteristic frequency of the wave. We denote the square amplitude by $a=\left(\tilde a^\alpha a_\alpha\right)^{1/2}$, wave vector by $l_\alpha={\cal S}_{;\alpha}$ and polarization vector by $m^\alpha=a^\alpha/a$. Now, let us expand both the wave and polarization vectors as
\begin{align}
    l^\alpha=& l^\alpha_0+\frac{l^\alpha_1}{\omega}+\frac{l^\alpha_2}{\omega^2}+ ...,\\
    m^\alpha=& m^\alpha_0+\frac{m^\alpha_1}{\omega}+\frac{m^\alpha_2}{\omega^2}+ ....
\end{align}
The necessity of expanding both the wave vector and amplitude comes from the fact that higher-order phase factors like ${\cal S}_1(\lambda)$ can not be absorbed into the leading-order amplitude $m^\alpha_0$ by transformation $m^\alpha \to e^{i {\cal S}_1 (\lambda)/\omega} m^\alpha$. This is because geometric optics require using a parallel propagated null tetrad, which constrains its transformation freedom $m^\alpha \to e^{i {\cal S}_1 (\lambda)/\omega} m^\alpha$ providing an additional restriction
\begin{equation}
    \frac{d {\cal S}_1 (\lambda)}{d\lambda}= 0. \label{gt11}
\end{equation}
Next, we substitute this vector potential onto the Lorentz gauge condition for electromagnetic waves, from which we obtain
\begin{equation}
    l_0^\alpha m_{0 \alpha}+\frac{1}{\omega}\left(l_0^\alpha m_{1 \alpha}+l_1^\alpha m_{0 \alpha}- i \left(\frac{a_{;\alpha}}{a}m_0^\alpha+m^\alpha_{0;\alpha}\right)\right)=0.\label{lc11}
\end{equation}
Again, substituting the vector potential into the source-free electromagnetic wave equation, we obtain
\begin{multline}
    j^\alpha \eqdef m_0^\alpha l_{0\beta} l_0^\beta+\frac{1}{\omega}\bigg(m_1^\alpha l_{0\beta} l_0^\beta+ 2 m_0^\alpha l_{1\beta} l_0^\beta\\
    -i \left(m_0^\alpha l^\beta_{0;\beta}+2 m^\alpha_{0;\beta}l_0^\beta+2 \frac{a_{;\beta}} {a}m_0^\alpha l_0^\beta \right) \bigg)=0,\label{we12}
\end{multline}
up to the subleading order in $\omega$. Let us now calculate an identically vanishing quantity, $\tilde m_{0\alpha}j^\alpha+ m_{0\alpha} \tilde j^\alpha$
\begin{equation}
     l_{0\beta} l_0^\beta+\frac{2}{\omega} \left(l_{1 \beta}-b_\beta \right) l_0^\beta =0.\label{dr13}
\end{equation}
This relation can be considered as the generalization of the dispersion relation of geometric optics. For simplification, we have applied $\tilde m_0^\alpha m_{0 \alpha}=1$ and substituted
\begin{equation}
    \frac{i}{2}\left(\tilde m^\alpha m_{\alpha;\beta}- m^\alpha \tilde m_{\alpha; \beta} \right)= i \tilde m^\alpha m_{\alpha;\beta} \defeq b_\beta. \label{b14}
\end{equation}

\section{Equations of spin optics}\label{soD}

To find the propagation equation of spin optics, let us start with the generalized dispersion relation~\eqref{dr13} and express it as
\begin{equation}
    \frac{1}{2} g_{\alpha\beta}l_0^\alpha l_0^\beta+\frac{1}{\omega} g_{\alpha\beta} \left(l_1^ \alpha-b^\alpha \right) l_0^\beta =0. \label{hj47}
\end{equation}
This relation can be viewed as the Hamilton-Jacobi equation for the subleading order phase function ${\cal S}$, where ${\cal S}_{;\alpha}= l_{0\alpha}+ l_{1\alpha}/\omega$. The Hamiltonian function on the cotangent bundle $T^* M$, associated with the Hamilton-Jacobi equation, is
\begin{multline}
    H(x,l)= \frac{1}{2} g^{\alpha\beta}l_{0\alpha} l_{0\beta}+\frac{1}{\omega} g^{\alpha\beta} \left(l_{1 \alpha}-b_\alpha \right) l_{0\beta}\\
    = \frac{1}{2\omega^2} g^{\alpha\beta} \left(\omega l_{0 \alpha} + l_{1\alpha}- b_\alpha\right) \left(\omega l_{0\beta} + l_{1\beta}- b_\beta\right).\label{he65}
\end{multline}
Hamilton's equations of motion are
\begin{equation}
    \frac{d x^\alpha}{d \lambda}= \frac{\partial H}{\partial l_{\alpha}}= g^{\alpha\beta} \left(l_ \beta- \frac{b_\beta}{\omega}\right), \label{tr64}
\end{equation}
and
\begin{equation}
    \frac{d l_\alpha}{d \lambda}= -\frac{\partial H}{\partial x^\alpha}=  \frac{1}{2} \dot x^\mu \dot x^\nu \frac{\partial g_{\mu\nu}}{\partial x^\alpha}+ \frac{1}{\omega} g^{\mu\nu} \dot x_\nu \frac{\partial b_\mu}{\partial x^\alpha}, \label{eom50}
\end{equation}
where we have used Eq.~\eqref{tr64} in obtaining this. Now, we could write the action whose solution from the variation principle corresponds to the solution of the Hamilton-Jacobi equation~\eqref{hj47} (see Ref.~\cite{pkd2})
\begin{multline}
    {\cal S} (x,l)= \int_{\lambda} (\dot x^\alpha l_\alpha- H(x,l)) d\lambda\\
    = \frac{1}{2} \int \dot x^\alpha \dot x_\alpha d\lambda + \frac{1}{\omega} \int b_\alpha \dot x^\alpha d\lambda, \label{a51}
\end{multline}
where Eqs.~\eqref{he65} and \eqref{tr64} are used for simplification. The first term is the optical path length, and the second term resembles the Berry connection of optics. Either the variational principle $\delta{\cal S}=0$ or simplification of Hamilton's equation of motion~\eqref{eom50} yields the same equation~\cite{pkd2}
\begin{equation}
    \frac{D^2 x_\mu}{D\lambda^2}+ \frac{1}{\omega} \left( b_{\mu;\nu}- b_{\nu;\mu}\right) \dot x^\nu= 0. \label{vp62}
\end{equation}
Further simplification of $b_{\beta;\alpha}- b_{\alpha;\beta} \defeq k_{\alpha\beta}$ gives
\begin{equation}
    k_{\alpha\beta}=-i R_{\alpha\beta\mu\nu}m^\mu \tilde m^\nu+ i \left(\tilde m^\nu_{;\alpha} m_{\nu;\beta}- \tilde m^\nu_{;\beta} m_{\nu;\alpha}\right). \label{k58}
\end{equation}
After substituting this back into Eq.~\eqref{vp62}, we obtain
\begin{equation}
    \frac{D^2 x^\alpha}{D\lambda^2}=-\frac{i}{\omega} R^\alpha_{~\beta\mu\nu}m^\mu \tilde m^\nu l_0^\beta \approx -\frac{i}{\omega} R^\alpha_{~\beta\mu\nu} l_0^\beta m_0^\mu \tilde m_0^\nu,\label{te43}
\end{equation}
which is the propagation Eq.~\eqref{te56}.

 \end{document}